\documentclass[epj]{svjour}
\usepackage{epsfig}
\usepackage{color}
\usepackage{amsmath}

\newcommand{\nn}{{\nonumber}\\}
\def\Id{{\rm 1\kern-.3em I}}
\newcommand{\id}{\mbox{{\small 1}\hspace{-0.37em}1}}
  \newbox\charbox
  \newbox\slabox
  \def\s#1{{      % Feynman slash
        \setbox\charbox=\hbox{$#1$}
        \setbox\slabox=\hbox{$/$}
        \dimen\charbox=\ht\slabox
        \advance\dimen\charbox by -\dp\slabox
        \advance\dimen\charbox by -\ht\charbox
        \advance\dimen\charbox by \dp\charbox
        \divide\dimen\charbox by 2
        \raise-\dimen\charbox\hbox to \wd\charbox{\hss/\hss}
        \llap{$#1$}
        }}
\begin{document}

\title{Charmed baryons in a relativistic quark model}
\author{Sascha Migura \and Dirk Merten \and Bernard Metsch \and Herbert-R.~Petry}
%\author{First author\inst{1} \and Second author\inst{2}% etc
\institute{Helmholtz-Institut f\"ur Strahlen- und Kernphysik, Nu{\ss}allee 14--16, 53115 Bonn, Germany \\ \email{migura@itkp.uni-bonn.de}}
%\institute{Insert the first address here \and the second here}
%
\date{Received: 16 February 2006 / Revised version: 13 April 2006}
\abstract{
We calculate mass spectra of charmed baryons within a relativistically covariant quark model based on the Bethe-Salpeter-equation in instantaneous approximation. Interactions are given by a linearly rising three-body confinement potential and a flavor dependent two-body force derived from QCD instanton effects. This model has already been successfully applied to the calculation of light flavor baryon spectra and is now extended to heavy baryons. Within the same framework we compare the results to those obtained with the more conventional one-gluon-exchange potential.
\PACS{
{11.10.St}{Bound and unstable states; Bethe-Salpeter equations}\and 
{12.39.Ki}{Relativistic quark model}\and
{14.20.Lq}{Charmed baryons}
     } % end of PACS codes
} %end of abstract
\maketitle

\section{Introduction}
This paper is part of a series studying properties of hadron resonances within a unified constituent quark model. The model is based on the Bethe-Salpeter-equation in instantaneous approximation and hence is relativistically covariant by construction. The present work extends the investigations of baryon spectra from previous papers on the non-strange \cite{Loring:2001kx} and the strange \cite{Loring:2001ky} sector to charmed baryons. We have found that our Bethe-Salpeter-model is able to describe not only the Regge-trajectories of light flavor baryon spectra but also their hyperfine structures with only seven free parameters. On this basis we have now calculated mass spectra of single, double and triple charm baryons.

A similar extension from light to heavy flavor ha\-drons within the Bethe-Salpeter-model has already been successfully performed for mesons, see \cite{Merten:2001er} for the results.

In the case of charmless baryons we used a three-quark interaction given by a linearly rising confinement potential equipped with a suitable spinoral structure that has been fixed by light flavored baryons \cite{Loring:2001kx}. This interaction remains completely unchanged in the present work. 

As we will show this ansatz also yields a satisfactory description of charmed baryons and hence the confining potential is flavor independent and universal.

The residual interaction we use is a phenomenological extension of 't Hooft's instanton-induced force to charm quarks (see \cite{'tHooft:1976fv,Shifman:1979uw} for details on instantons). Phenomenologically here means that the standard derivation is not justified for massive quarks and applicable only for nearly massless quarks. The couplings that enter into this two-body 't Hooft-con\-tri\-bution are for light quarks fixed phenomenologically by the non-strange and the strange bary\-on spectra and are left unchanged. Two additional couplings enter when including charm quarks and are treat\-ed as free parameters. The constituent quark mass of the charm\-ed quark represents another free parameter as well.

These three new parameters are fixed by fitting the experimental single charm baryon spectrum to which we compare our calculations in detail.

Acting between a heavy and a light quark, the one-gluon-exchange is an alternative residual two-quark interaction and is investigated within the same framework. We thus will compare the influence of the one-gluon-exchange on the charmed baryon spectrum to the effects of the extended instanton force. 

With these parameters fixed we go a step further and calculate the masses of the lowest double and triple charm baryons and compare the results to the very sparse experimental data. A number of predictions is added.

The paper is organized as follows: Section \ref{sec:BSM} very briefly recapitulates the Bethe-Salpeter-model and shows the ingredients and basic equations. For more details on the theory of the Bethe-Salpeter-model we refer to \cite{Loring:2001kv}. Section \ref{sec:interaction} shortly explains the interactions we use, especially the phenomenological extension of 't Hooft's force to charm quarks. Section \ref{sec:results} summarizes all model parameters and our results for the spectrum of charmed baryons. These results are discussed in detail in comparison to the experimental data. We display the results for the one-gluon-exchange as a residual interaction as well and discuss calculated mass spectra for double and triple charm baryons before concluding in sec.~\ref{sec:conclusion}. 

\vspace{20pt}

\section{Bethe-Salpeter-model} \label{sec:BSM} 

\subsection{Bound states}
Our formally covariant constituent quark model is based on the Bethe-Salpeter-amplitude $\chi$ defined via quark field operators $\Psi_{a_i}(x_i)$ by
\begin{eqnarray}
\chi_{\bar P\,a^{}_1 a^{}_2 a^{}_3}(x^{}_1,x^{}_2,x^{}_3)
=\langle0|T\Psi^{}_{a^{}_{1}}(x^{}_{1})\Psi^{}_{a^{}_{2}}(x^{}_{2})
\Psi^{}_{a^{}_{3}}(x^{}_{3})|\bar P\rangle \label{eqn:BSA}
\end{eqnarray}
where $\bar P$ denotes the four-momentum of the on-shell bound state and $T$ the time ordering operator. The Bethe-Sal\-pe\-ter-amplitude describes the baryonic bound states and replaces the usual quantum mechanical wave function in non-relativistic quark models. Due to translational invariance we switch from now on to relative coordinates.

\begin{figure}[t]
\begin{center}
\input{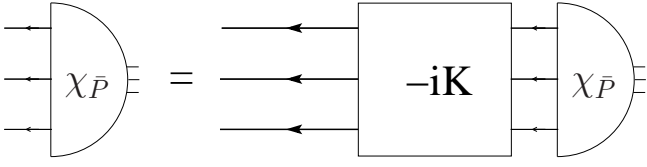}
\caption{Graphical illustration of the three-quark Bethe-Salpeter-equation (\ref{eqn:BSE}). Thick arrows indicate the full quark propagators $S^i_F$. $K$ contains the three-body confinement and the two-body residual interactions.}
\label{fig:BSE}
\end{center}
\end{figure}

The Fourier-transform of the Bethe-Salpeter-amplitude is determined by the Bethe-Salpeter-equation
\begin{eqnarray}
&& \chi_{\bar P\,a_1 a_2 a_3}(p_\xi,p_\eta)=
S^1_{F\,a_1 a'_1}\left(\textstyle \frac{1}{3}\bar P+p_{\xi}+\frac{1}{2}p_{\eta}\right)\, \nonumber \\
&& \,\times S^2_{F\,a_2 a'_2}\left(\textstyle \frac{1}{3}\bar P-p_{\xi}+\frac{1}{2}p_{\eta}\right)\,
S^3_{F\,a_3 a'_3}\left(\textstyle \frac{1}{3}\bar P-p_{\eta}\right)\nonumber \\
&& \,\times\, (-\mathrm{i})\,
\int\frac{\mathrm{d}^4 p_\xi'}{(2\pi)^4}\,\frac{\mathrm{d}^4 p_\eta'}{(2\pi)^4}\,
K^{}_{\bar P\, a'_1 a'_2 a'_3;\, a''_1 a''_2 a''_3}
(p_\xi,p_\eta;\, p_\xi',p_\eta')\nonumber \\
&& \quad\quad\quad\quad\quad\quad\quad\quad\quad\quad\quad\quad\quad\times\chi_{\bar P\, a''_1 a''_2 a''_3}(p_\xi',p_\eta') \label{eqn:BSE}
\end{eqnarray}
where full quark propagators are denoted by $S^i_F$. $K$ stands for the irreducible interaction kernel that contains both irreducible two- and three-quark interaction kernels, $K^{(2)}$ and $K^{(3)}$ respectively. See fig.~\ref{fig:BSE} for a graphical illustration.

\subsection{Approximations}
Solving eq.~(\ref{eqn:BSE}) rigorously is presently not possible: The propagators and the interaction kernels are sums of an infinite number of Feynman-graphs and are unknown functions within QCD. Furthermore, the dependence of the Bethe-Salpeter-equation (\ref{eqn:BSE}) on the relative energies leads to a complicated analytic pole structure.

We therefore replace the full quark propagator $S^i_F$ by the usual fermion propagator
\begin{eqnarray}
S^{i}_F(p_i)=\frac{\mathrm{i}}{\s p_i-m_i+\mathrm{i}\epsilon}
\end{eqnarray}
where $m_i$ denotes the effective constituent quark mass and enters as a free parameter. As a second approximation we choose the interaction kernels to be instantaneous in the rest frame of the baryon with mass $M$. This means that these do not depend on their relative energies, \textit{i.e.}
\begin{eqnarray}
K^{(3)}_{P}(p^{}_{\xi},p^{}_{\eta},p'_{\xi},p'_{\eta})\Bigg|_{P=(M,\vec
0)}=V^{(3)}(\vec p^{}_{\xi},\vec p^{}_{\eta},\vec
p_{\xi}\,\!\!', \vec p_{\eta}\,\!\!') 
\end{eqnarray}
and
\begin{eqnarray}
K^{(2)}_{\frac{2}{3}P+p_{\eta_k}}(p^{}_{\xi_k},p'_{\xi_k})\Bigg|_{P=(M,\vec 0)}=V^{(2)}(\vec p^{}_{\xi_k},\vec p_{\xi_k}\,\!\!') 
\end{eqnarray}
where we have suppressed the multi-indices. The index $k$ numbers the three possible quark pairs.

\subsection{Salpeter-equation}
When genuine two-quark kernels $K^{(2)}$ are involved, special difficulties arise if one now wants to eliminate the relative energies in eq.~(\ref{eqn:BSE}). Because of the third non-interacting spectator quark there remains a relative energy dependence. 

However, we can eliminate the relative energies in eq. (\ref{eqn:BSE}) by introducing the projected Salpeter-amplitude
\begin{eqnarray}
\Phi^{\Lambda}_M(\vec p^{}_\xi,\vec p^{}_\eta)=\Lambda_+(\vec p^{}_\xi,\vec p^{}_\eta)\,\Phi^{}_M(\vec p^{}_\xi,\vec p^{}_\eta)
\end{eqnarray}
where we use the usual Salpeter-amplitude
\begin{eqnarray}
\Phi^{}_M(\vec p^{}_\xi,\vec p^{}_\eta)=\int
\frac{\mathrm{d}p^0_\xi}{2\pi}\frac{\mathrm{d}p^0_\eta}{2\pi}\,\chi^{}_M(p^{}_\xi,p^{}_\eta).
\label{DefSalAmpl}
\end{eqnarray}
and the Salpeter-projector
\begin{eqnarray}
\Lambda_{\pm}(\vec p_\xi,\vec p_\eta)&=&\Lambda^+(\vec
p_1)\otimes\Lambda^+(\vec p_2)\otimes\Lambda^+(\vec
p_3)\nonumber\\
&&\pm\Lambda^-(\vec p_1)\otimes\Lambda^-(\vec
p_2)\otimes\Lambda^-(\vec p_3).
\end{eqnarray}
The projection operator reads
\begin{eqnarray}
\Lambda^\pm(\vec p_{i})=\sum_f\Lambda^\pm_{m_f}(\vec p_{i})\otimes\mathcal{P}^\mathcal{F}_f
\end{eqnarray}
where
\begin{eqnarray}
\Lambda^\pm_{m_f}(\vec p_{i})=\frac{\omega_{m_f}(\vec p_{i})\pm
H_{m_f}(\vec p_i)}{2\omega_{m_f}(\vec p_{i})}
\end{eqnarray}
and $\mathcal{P}^\mathcal{F}_f=|f\rangle\langle f|$ is the flavor projector  that assigns the correct quark masses with $f$ running over all flavors. The one-quark energy is
\begin{eqnarray}
\omega_{m_f}(\vec p_{i})=\sqrt{\vec p_{i}\;\!\!^{2}+m^2_{f}}
\end{eqnarray}
and the Dirac-Hamilton-operator 
\begin{eqnarray}
H_{m_f}(\vec p_i)=\gamma^0(\vec\gamma\cdot\vec p_i+m_{f}).
\end{eqnarray}

As shown in detail in \cite{Loring:2001kv}, eq.~(\ref{eqn:BSE}) can now be reduced in Born-approximation to an eigenvalue problem of the form
\begin{equation}
\mathcal{H}\Phi_M^\Lambda = M\,\Phi_M^\Lambda
\label{eqn:SE}
\end{equation}
for the projected Salpeter-amplitude, the eigenvalues being the baryon masses $M$. The Salpeter-Ha\-mil\-tonian $\mathcal{H}$ explicitly reads
\begin{eqnarray}
\label{eqn:SE_Hamiltonian}
&&\left[\mathcal{H}\Phi_M^\Lambda\right]({\bf p_\xi}, {\bf p_\eta})
= 
\mathcal{H}_0({\bf p_\xi}, {\bf p_\eta})\;\Phi_M^\Lambda({\bf p_\xi}, {\bf p_\eta})\nonumber\\
&&+\Lambda_{+}(\vec p_\xi,\vec p_\eta)\,\gamma^0\otimes\gamma^0\otimes\gamma^0\;
\int 
\frac{\textrm{d}^3 p_\xi'}{(2\pi)^3}\;
\frac{\textrm{d}^3 p_\eta'}{(2\pi)^3}\;\nonumber\\
&&\hspace{80pt} V^{(3)}({\bf p_\xi},{\bf p_\eta};\;{\bf p_\xi'},{\bf
  p_\eta'})\;
\Phi_M^\Lambda({\bf p_\xi'}, {\bf p_\eta'})\nn
&&+\Lambda_{-}(\vec p_\xi,\vec p_\eta)\,\gamma^0\otimes\gamma^0\otimes\Id\;
\int \frac{\textrm{d}^3 p_\xi'}{(2\pi)^3}\;\nonumber\\
&&\hspace{80pt} V^{(2)}({\bf p_\xi},{\bf p_\xi'})\otimes\Id\;
\Phi_M^\Lambda({\bf p_\xi'}, {\bf p_\eta})\\
&&+ \textrm{terms with interacting quark pairs (23) and (31)} \nonumber
\end{eqnarray}
where
\begin{equation}
\mathcal{H}_0=H(\vec p_1)\otimes\Id\otimes\Id+\Id\otimes H(\vec p_2)\otimes\Id+\Id\otimes\Id\otimes H(\vec p_3)
\end{equation}
denotes the free three-quark Hamiltonian with
\begin{eqnarray}
H(\vec p_{i})=\sum_f H_{m_f}(\vec p_{i})\otimes\mathcal{P}^\mathcal{F}_f.
\end{eqnarray}

Equation (\ref{eqn:SE}) is call\-ed the Salpeter-equation and can be solved by standard techniques. This procedure has been already successfully applied to the calculation of light bary\-ons, see \cite{Loring:2001kx,Loring:2001ky,Loring:2001kv}. The spectra describe very well the experimental Regge-trajectories and hyperfine structures with a very limited number of free parameters.

\section{Interactions} \label{sec:interaction}
We will now specify the interaction kernels $V^{(3)}$ and $V^{(2)}$ which enter into eq.~(\ref{eqn:SE_Hamiltonian}). We assume confinement to be a pure three-quark potential; the instanton-induced force and alternatively one-gluon-exchange are given by two-quark residual interactions.

\subsection{Confinement}
Quarks have the property to be asymptotically free at short distances but to interact strongly for large distances in the low and intermediate energy region. It is still not possible to derive their confinement analytically from QCD but it is clearly visible in lattice calculations. We thus parameterize the confinement potential phenomenologically. As shown in \cite{Loring:2001kx,Loring:2001ky} the instantaneous ansatz
\begin{eqnarray}
&& V^{(3)}(x^{}_1,x^{}_2,x^{}_3;x'_1,x'_2,x'_3)
 =V^{(3)}_{\mathrm{conf}}(\vec x^{}_1,\vec x^{}_2,\vec
x^{}_3)\delta(x^0_1-x^0_2) \nonumber \\
&& \times\delta(x^0_2-x^0_3) \delta^{(4)}(x^{}_1-x'_1)\delta^{(4)}(x^{}_2-x'_2)\delta^{(4)}(x^{}_3-x'_3)
\end{eqnarray}
given in coordinate space with
\begin{eqnarray}
&&V^{(3)}_{\mathrm{conf}}(\vec x^{}_1,\vec x^{}_2,\vec x^{}_3)
=3a\cdot\frac{1}{4}(\id\otimes\id\otimes\id+
\gamma^0\otimes\gamma^0\otimes\id \nonumber\\ &&\hspace{10pt}+\textrm{cycl.~perm.})
+b\sum_{i<j}|\vec x_i-\vec x_j|\cdot\frac{1}{2}(-\id\otimes\id\otimes\id
\nonumber \\ &&\hspace{85pt}+\gamma^0\otimes\gamma^0\otimes\id+\textrm{cycl.~perm.})
\end{eqnarray}
leads to satisfactory results when describing light flavor baryon spectra. The offset $a$ and the slope $b$ enter as free parameters and are fixed by the light flavor baryon spectrum. We do not change these confinement parameters and thus assume that confinement is indeed universal and flavor independent.

\subsection{'t Hooft's potential}
Instantons are special solutions of the classical Euclidean QCD Yang-Mills-equations. They fall into topologically distinct homotopy classes and can in Minkowski-space be interpreted as tunneling events between distinct vacua.

As shown for the first time in \cite{'tHooft:1976fv} instantons can contribute to quark interactions. In particular, instantons lead to an effective two-quark interaction which for quark pairs in baryons reads
\begin{eqnarray}
&&V^{(2)}(x^{}_1,x^{}_2;x'_1,x'_2)=V^{(2)}_{\mbox{\scriptsize
't Hooft}} (\vec x^{}_1-\vec x^{}_2)
\delta(x^0_1-x^0_2)\nonumber\\
&&\hspace{95pt}\delta^{(4)}(x^{}_1-x'_1)\delta^{(4)}(x^{}_2-x'_2)
\end{eqnarray}
with
\begin{eqnarray}
&&V^{(2)}_{\mbox{\scriptsize 't Hooft}}(\vec x)
=-4v_{\mathrm{reg}}(\vec x)
\times(\id\otimes\id+\gamma^5\otimes\gamma^5)
\mathcal{P}^{\mathcal{D}}_{S_{12}=0}\nonumber\\&&
\otimes(g_{nn}\mathcal{P}^{\mathcal{F}}_{\mathcal{A}}(nn)
+g_{ns}\mathcal{P}^{\mathcal{F}}_{\mathcal{A}}(ns)
+g_{nc}\mathcal{P}^{\mathcal{F}}_{\mathcal{A}}(nc)
+g_{sc}\mathcal{P}^{\mathcal{F}}_{\mathcal{A}}(sc))\nonumber\\
\label{eqn:instantonact}
\end{eqnarray}
where $\mathcal{P}^{\mathcal{D}}_{S_{12}=0}$ is the projector onto spin-singlet states and $\mathcal{P}^{\mathcal{F}}_{\mathcal{A}}(f_1f_2)$ is the projector onto flavor-antisymmetric quark pairs with flavors $f_1$ and $f_2$. The couplings $g_{f_1f_2}$ are given by integrals over instanton densities which are basically unknown. We treat them here as free parameters.

Note that we have ad hoc extended 't Hooft's force to charmed quarks by introducing two additional couplings $g_{nc}$ and $g_{sc}$ and flavor projectors $\mathcal{P}^{\mathcal{F}}_{\mathcal{A}}(nc)$ and $\mathcal{P}^{\mathcal{F}}_{\mathcal{A}}(sc)$ that operate exactly like the terms for the light-flavored quarks.

Due to the non-perturbative approach we also had to regularize 't Hooft's potential which is essentially a contact interaction that we have replaced by a Gaussian function of the form      
\begin{eqnarray}
v_{\mathrm{reg}}(\vec x)=\frac{1}{\lambda^3\pi^{\frac{3}{2}}} e^{-\frac{|\vec x|^2}{\lambda^2}}.
\end{eqnarray}
The effective range parameter $\lambda$ enters as an additional free parameter assumed to be flavor independent.

\subsection{One-gluon-exchange}
For comparison we investigate also the one-gluon-exchange as a residual interaction by assuming a gluon propagator
\begin{eqnarray}
\gamma^\mu D_{\mu\nu}\gamma^\nu
=
4\pi
\left(\frac{\gamma^0\otimes\gamma^0}{|{\vec q|^2 }}
+
\frac{ 
\boldsymbol{\gamma}\cdot\!\otimes\;\boldsymbol{\gamma} 
- 
(\boldsymbol{\gamma}\cdot {\vec{\hat q}})\otimes(\boldsymbol{\gamma}\cdot {\vec{\hat q}})}
{q^2+\mathrm{i}\epsilon}
\right) \nonumber\\
\end{eqnarray}
in Coulomb-gauge with ${\vec{\hat q}}:= {\vec q}/|{\vec q}|$, see \cite{Murota:1982in}. The component $D_{00}(q)$, which describes the Coulomb-potential, is already instantaneous and we make the remaining components also instantaneous by setting $q^2$ equal to $-|\vec{q}|^2$. The two-quark one-gluon-exchange interaction then reads in coordinate space
\begin{eqnarray}
&&V^{(2)}(x^{}_1,x^{}_2;x'_1,x'_2)=V^{(2)}_{\mbox{\scriptsize
OGE}} (\vec x^{}_1-\vec x^{}_2)
\delta(x^0_1-x^0_2)\nonumber\\
&&\hspace{95pt}\delta^{(4)}(x^{}_1-x'_1)\delta^{(4)}(x^{}_2-x'_2)
\end{eqnarray}
with
\begin{eqnarray}
&&V^{(2)}_{\mathrm{OGE}}({\bf x})
=-\frac{2}{3}\frac{\alpha_s}{|{\bf x}|}
\Big(\gamma^0\otimes\gamma^0 - 
\frac{1}{2}\boldsymbol{\gamma}\cdot\!\otimes\;\boldsymbol{\gamma}
\nonumber\\&& \hspace{100pt}
- \frac{1}{2}(\boldsymbol{\gamma}\cdot {\bf \hat x})\otimes(\boldsymbol{\gamma}\cdot {\bf \hat x})
\Big) \label{eqn:OGEpot}
\end{eqnarray} 
where ${\bf \hat x}:= {\bf x}/|{\bf x}|$, see again \cite{Murota:1982in}. We treat the strong coupling constant $\alpha_s$ as a fit parameter.

\section{Results and discussion} \label{sec:results}

\subsection{General remarks} 
The baryons we consider can be classified into flavor SU(4) multiplets. The subsets of defined symmetry under permutation of the three quarks are obtained as irreducible representations of the threefold product of the fundamental representation of SU(4), \textit{i.e.}
\begin{eqnarray}
4\otimes 4\otimes 4=20_{\mathrm{S}}\oplus 20_{\mathrm{M}_\mathrm{S}}\oplus 20_{\mathrm{M}_\mathrm{A}}\oplus\bar{4}_{\mathrm{A}}.
\end{eqnarray}
The indices stand for total and mixed symmetry and anti-symmetry. We are mainly interested in single charm bary\-ons which reduces the flavor space to a symmetric 6-plet containing $\Sigma_c$, $\Xi_c$ and $\Omega_c$, a mixed 6-plet containing $\Sigma_c$, $\Xi'_c$ and $\Omega_c$, a mixed and an anti-symmetric $\bar 3$-plet containing both $\Lambda_c$ and $\Xi_c$. Note that we have exact SU(2)-isospin symmetry in the Bethe-Salpeter-model.

The physical ground states $\Xi'_c\frac{1}{2}^+$ and $\Xi_c\frac{1}{2}^+$ are linear combinations of states belonging to mixed multiplets. This will be further investigated in the discussion.

The Particle Data Group Collaboration lists fourteen detected charmed baryons \cite{Eidelman:2004wy}, thirteen single and only one double charm state. No spin or parity quantum numbers have actually been measured yet but are plausible quark model assignments. There are no quantum numbers known for the one two- and the two one-star states, among them the double charm candidate. 

The only ground state of the single charm multiplets that has not been detected yet is the $\Omega_c\frac{3}{2}^+$ belonging to the symmetric 6-plet.

\renewcommand{\arraystretch}{1.25}

\subsection{Parameters}
Before we present the results for the spectra of charmed baryons calculated within the Bethe-Salpeter-model, we show in table~\ref{tab:parameters} the explicit numbers of the free parameters that enter into the calculations.
\begin{center}
\begin{table}[t]
\begin{center}
\begin{tabular}{|c|c||c|c|}
\hline
$m_n$ & 330\,MeV & $g_{nn}$ &136.0\,MeV\,fm$^3$ \\
$m_s$ & 670\,MeV & $g_{ns}$ &94.0\,MeV\,fm$^3$  \\
$m_c$ & 1950\,MeV& $g_{nc}$ & 33.3\,MeV\,fm$^3$\\
$a$     & -744\,MeV& $g_{sc}$ & 11.0\,MeV\,fm$^3$\\
$b$ & 470\,MeV\,fm$^{-1}$&$\lambda$&0.4 fm\\
\hline
\end{tabular}
\end{center}
\caption{All ten free parameters of the Bethe-Salpeter-model. In the left columns are the constituent quark masses and the confinement parameters. On the right are the couplings and the effective range of the 't Hooft-interaction.}
\label{tab:parameters}
\end{table}
\end{center}
We have added three additional free parameters, namely the constituent quark mass $m_c$ and the couplings $g_{nc}$ and $g_{sc}$, to the seven parameters already fixed by the light-flavored baryon spectrum.

There is no contribution of 't Hooft's force for flavor symmetric states, such as $\Delta$-resonances. So the off-set parameter $a$ and the slope $b$ of the confinement potential and the non-strange quark mass $m_n$ have been fixed by the well measured positive-parity $\Delta$-Regge trajectory alone. The strange quark mass $m_s$ is fitted to the decuplet hyperons that are not affected by the 't Hooft-potential either. The coupling $g_{nn}$ and the effective range $\lambda$ are fixed by the $\Delta-N$ mass splitting whereas the coupling $g_{ns}$ is determined by the experimental hyperfine splitting between octet and decuplet hyperon ground states. 

The new free parameters $m_c$, $g_{nc}$ and $g_{sc}$ are simultaneously fitted to the eleven experimentally known three- and four-star single charm baryons listed in \cite{Eidelman:2004wy}.

\subsection{Spectrum of charmed baryons} \label{sec:Spectra}
\begin{figure*}[t]
\begin{center}
\input{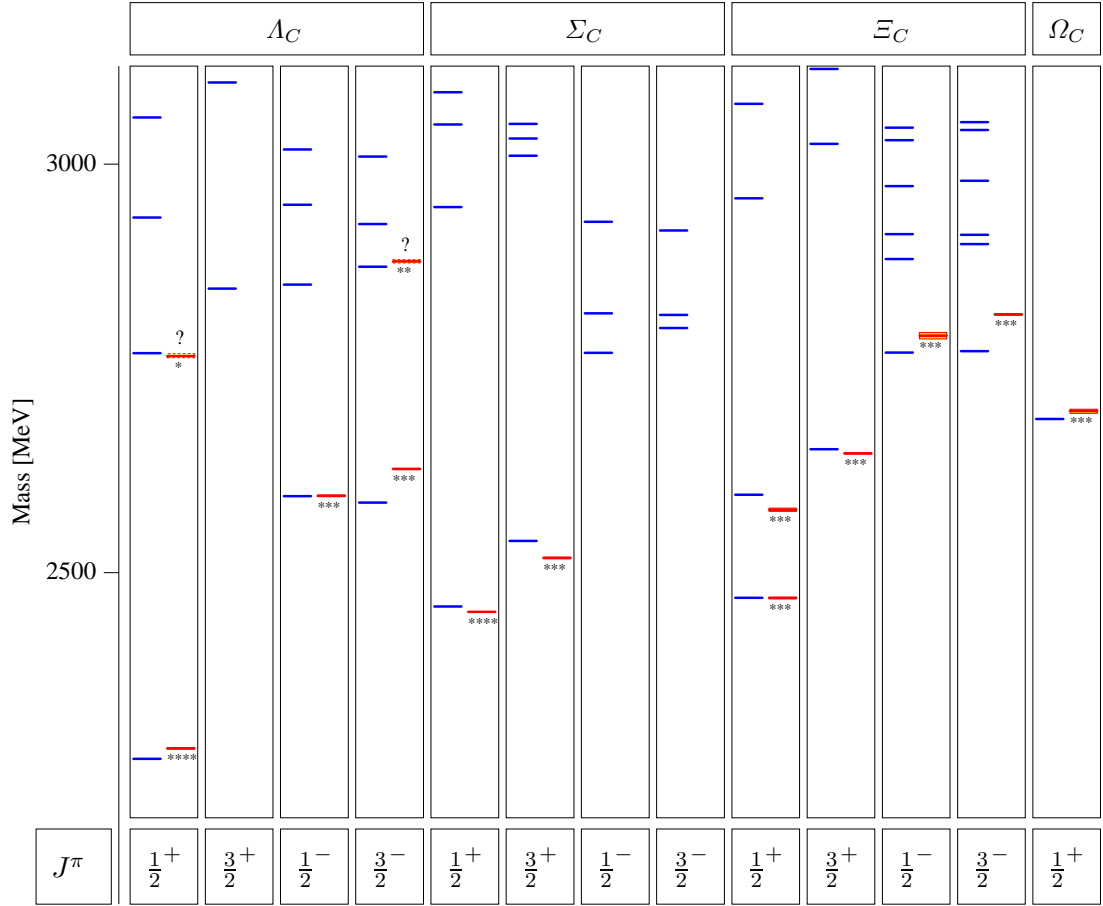}
\caption{The single charm baryon spectrum calculated in the Bethe-Salpeter-model using an appropriate three-quark confinement potential and 't Hooft's force between all possible flavor combinations (on the left side of each column) in comparison to the experimental masses from \cite{Eidelman:2004wy} (on the right side where the status is indicated by stars, the mass uncertainty by a shaded box and the lack of knowledge on quantum numbers by a question mark) for various total angular momentum $J$ and parity $\pi$ assignments given by our model.}
\label{fig:charmed}
\end{center}
\end{figure*}

Table \ref{tab:singlycharmedmasses} compares the numbers of all thirteen known experimental masses to our computed values for single charm baryonic states using 't Hooft's force between all possible flavor combinations as given in eq.~(\ref{eqn:instantonact}). 
\begin{center}
\begin{table}[t]
\begin{center}
\begin{tabular}{|c|c|c|c|c|}
\hline
State&Rating&$J^\pi$&EXP&BSM\\
\hline
$\Lambda_c$      &****&$1/2^+$&$2285\pm 1$&2272\\
$\Lambda_c(2593)$&*** &$1/2^-$&$2594\pm 1$&2594\\ 
$\Lambda_c(2625)$&*** &$3/2^-$&$2627\pm 1$&2586\\
$\Lambda_c?\Sigma_c?(2765)$&*&$?^?$&$2765\pm 3$&2769\\
$\Lambda_c?\Sigma_c?(2880)$&**&$?^?$&$2881\pm 2$&2874\\ 
$\Sigma_c(2455)$ &****&$1/2^+$&$2452\pm 1$&2459\\ 
$\Sigma_c(2520)$ &*** &$3/2^+$&$2518\pm 2$&2539\\ 
$\Xi_c$          &*** &$1/2^+$&$2469\pm 1$&2469\\ 
$\Xi'_c$         &*** &$1/2^+$&$2577\pm 3$&2595\\ 
$\Xi_c(2645)$    &*** &$3/2^+$&$2646\pm 2$&2651\\ 
$\Xi_c(2790)$    &*** &$1/2^-$&$2790\pm 4$&2769\\ 
$\Xi_c(2815)$    &*** &$3/2^-$&$2816\pm 2$&2771\\ 
$\Omega_c$       &*** &$1/2^+$&$2698\pm 3$&2688\\ 
$\Omega_c$       &    &$3/2^+$&    &2721\\ 
\hline
\end{tabular}
\end{center}
\caption{Ratings, the experimental (EXP) positions with their error bars (in MeV) together with the quantum numbers total angular momentum $J$ and parity $\pi$ of all known single charm baryons \cite{Eidelman:2004wy}. These masses are compared to the values calculated within the Bethe-Salpeter-model (BSM) using an appropriate three-quark confinement potential and 't Hooft's force between all possible flavor combinations. For the one- and two-star resonances, for which the quantum numbers are not known, we have performed an assignment based on the spectrum given in fig.~\ref{fig:charmed}. We also predict the mass for the $\Omega_c\frac{3}{2}^+$ which has not been measured yet but completes the symmetric 6-plet.}
\label{tab:singlycharmedmasses}
\end{table}
\end{center}
Figure \ref{fig:charmed} displays not only the direct comparison between theoretical and known experimental resonances but also shows many states that we predict which have not been observed so far. We restrict ourselves to the energy region between 2200 and 3120\,MeV and to total spins not exceeding $\frac{3}{2}$.

Comparing the gross structure of the experimental and theoretical spectrum we find a very good overall agreement. We have a very clear one-to-one correlation between experimental and theoretical states: There is no experimental state that we do not predict and there is no low lying theoretical resonance that is not observed.

We assign the two-star resonance to the first excited $\Lambda_c\frac{3}{2}^-$-state and the one-star resonance to the first excited $\Lambda_c\frac{1}{2}^+$-state. These assignments are not mandatory. The two-star resonance could have the spin-parity combinations $\frac{3}{2}^+$ or $\frac{1}{2}^-$ just as well. The one-star resonance could also be the lowest lying $\Sigma_c\frac{1}{2}^-$-state.

The discrepancies between the experimental and theoretical masses are relatively small. Eleven of the thirteen predicted states have a relative mass deviation less than one percent. The remaining three baryons have a mass deviation that is approximately one and a half percent.

We underestimate the $\Lambda_c$ by 13\,MeV which is only 0.6\% of its total mass whereas the other four-star state $\Sigma_c(2455)$ is overestimated by 7\,MeV which is only 0.3\% of the total mass.

Both positions of the three-star resonances $\Lambda_c(2593)$ and $\Xi_c$ are exactly reproduced and the masses of the three-star states $\Sigma_c(2520)$, $\Xi_c'$, $\Xi_c(2645)$ and $\Omega_c$ differ only by 0.8\%, 0.7\%, 0.2\% and $-0.4$\% respectively. The first excited $\Omega_c\frac{1}{2}^+$-state is located at 3169\,MeV.

Not accounted for are the experimental splittings of 33\,MeV and 26\,MeV between the spin-parity $\frac{1}{2}^-$ and $\frac{3}{2}^-$ pairs $\Lambda_c(2593)$, $\Lambda_c(2625)$ and $\Xi_c(2790)$, $\Xi_c(2815)$. Where\-as these pairs each are almost degenerate in the Bethe-Sal\-pe\-ter-model they seem to be clearly separated experimentally. The theoretical splitting between the two $\Lambda_c$ resonances even has the wrong sign. This can also be found for the negative parity sector in the light flavor baryon spectra of the $\Delta$, $N$ and $\Lambda$ resonances, see \cite{Loring:2001kx,Loring:2001ky}. Consequently, we underestimate both masses of the $\frac{3}{2}^-$-states $\Lambda_c(2625)$ and $\Xi_c(2815)$ by 1.6\% which is still small but distinctly more than for all the other states. The theoretical splitting between the two $\Xi_c$ resonances has at least the correct sign but the theoretical mass for $\Xi_c(2790)$ is already too small by 0.8\%. 

The one- and two-star resonances are again very well described and differ only by 4\,MeV and $-7$\,MeV when compared to the experimental values.

In table \ref{tab:singlycharmedmasses} we have also given our prediction for the $\Omega_c\frac{3}{2}^+$ that is the only baryon of the symmetric 6-plet that has not been detected in experiment yet.

\subsection{Instanton effects between light quarks} \label{sec:instlight}

We shall now study in detail the effects of the 't Hooft-interaction between light quarks and a light and a charm quark by switching off the 't Hooft-couplings $g_{nn}$, $g_{ns}$, $g_{nc}$ and $g_{sc}$ and increasing gradually their strengths one after another to their final values given in table \ref{tab:parameters}. It has been shown in \cite{Loring:2001kx,Loring:2001ky} that instanton-induced effects indeed dominate the fine structure of the whole light-flavored baryon spectrum. We will show that this is also the case for charmed baryons.

\begin{figure*}[p]
\begin{center}
\input{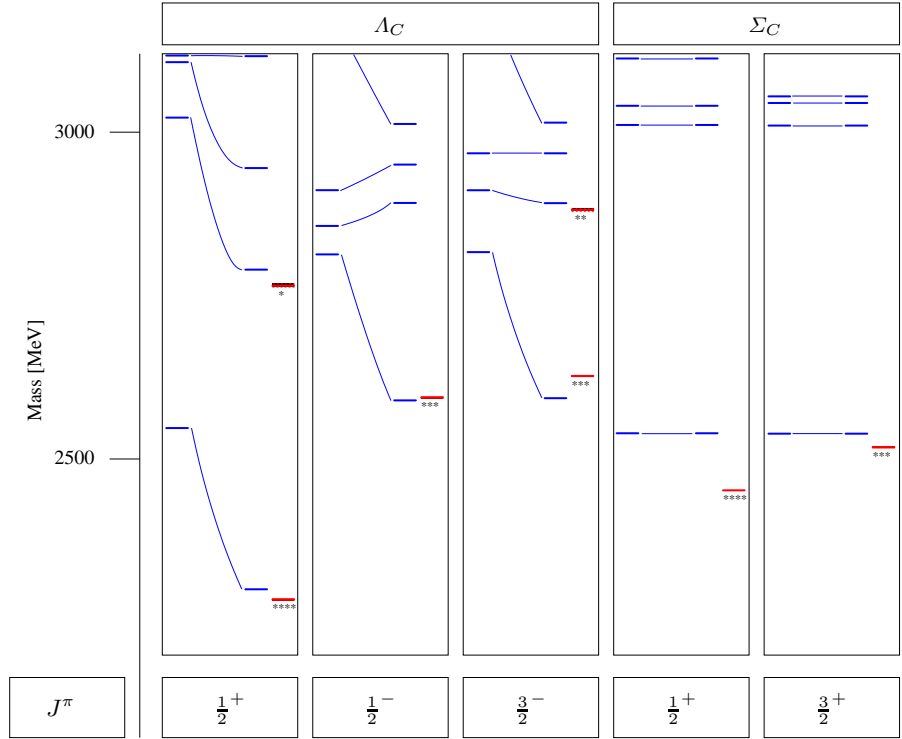}
\caption{The effects of the instanton-induced interaction between the light quarks on the theoretical single charm baryon resonances $\Lambda_c$ and $\Sigma_c$ with total angular momentum $J$ and parity $\pi$. The spectrum on the left in each column is determined by the confinement force alone. The following curves illustrate how the spectrum changes with increasing 't Hooft-coupling $g_{nn}$. The final value is shown in comparison with the experimental value shown on the right in each column.}
\label{fig:charmed_tHooft01}
\end{center}
\end{figure*}

\begin{figure*}[p]
\begin{center}
\input{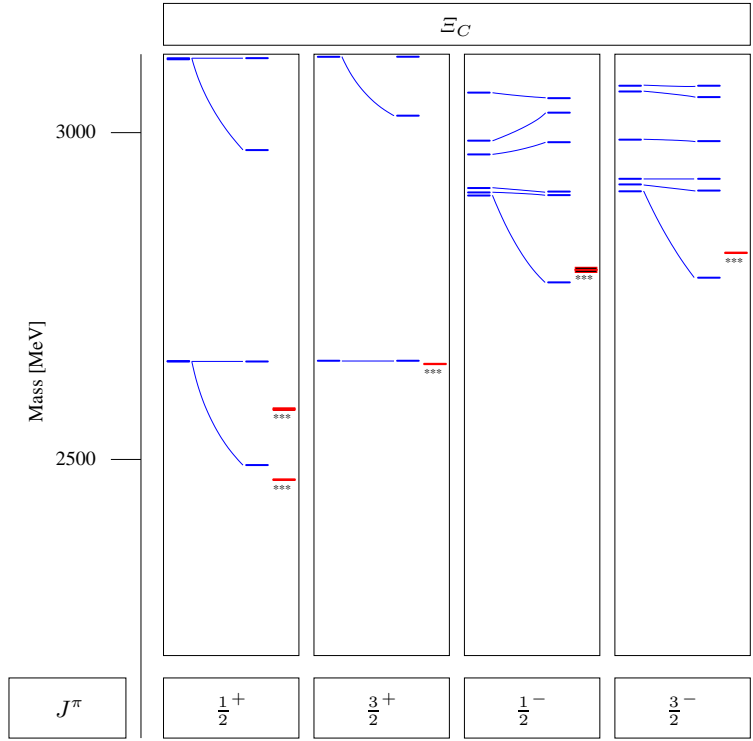}
\caption{The effects of the instanton-induced interaction between the light quarks on the resonances $\Xi_c$. The spectrum on the left in each column is determined by the confinement force alone. The curves illustrate how the spectrum changes with increasing 't Hooft-coupling $g_{ns}$. The final value is shown in comparison with the experimental value.}
\label{fig:charmed_tHooft02}
\end{center}
\end{figure*}

Figure \ref{fig:charmed_tHooft01} shows on the left in each column the theoretical single charm baryon spectrum of the $\Lambda_c$ and $\Sigma_c$ resonances determined by the confinement interaction alone which means that $g_{nn}=g_{nc}=0$. Due to their quark contents $nnc$ there never is a contribution from the coupling parameterized by $g_{ns}$.

We see that $\Sigma_c(2455)$, $\Sigma_c(2520)$ and $\Lambda_c(2880)$ are actually already well described by the confinement potential alone. It is true that the four-star state $\Sigma_c(2455)$ is with 87\,MeV clearly overestimated but we are now just interested in the gross picture.

\begin{center}
\begin{table}[t]
\begin{center}
\begin{tabular}{|c|rrrr|rr|}
\hline
Mass &
$\!\!{}^2 \bar 3_{\mathrm{M}}[\mathrm{S}]\!\!$&
$\!\!{}^2 \bar 3_{\mathrm{M}}[\mathrm{M}]\!\!$&
$\!\!{}^4 \bar 3_{\mathrm{M}}[\mathrm{M}]\!\!$&
$\!\!{}^2 \bar 3_{\mathrm{M}}[\mathrm{A}]\!\!$&
$\!\!{}^2 \bar 3_{\mathrm{A}}[\mathrm{M}]\!\!$&
$\!\!{}^4 \bar 3_{\mathrm{A}}[\mathrm{A}]\!\!$\\
\hline
\multicolumn{7}{|c|}{$J^\pi=1/2^+$}\\
\hline
     2272 & {\bf \underline{96.9}} &    1.9 &    0.0 &    0.1 &    0.9 &    0.1\\
     2769 & {\bf \underline{75.4}} &   11.7 &    0.1 &    0.2 &    12.1 &    0.4\\
     2935 &   20.6 &   21.3 &    0.0 &    1.6 & {\bf \underline{52.9}} &    3.4\\
     3057 &    1.2 & {\bf \underline{61.1}} &    3.7 &    1.0 &   32.6 &    0.1\\
\hline
\multicolumn{7}{|c|}{$J^\pi=3/2^+$}\\
\hline
     2848 & {\bf \underline{53.2}} &   24.6 &    0.3 &    0.1 &   21.6 &    0.2\\
     3100 &    2.2 &   24.4 &   18.7 &    0.9 & {\bf \underline{50.7}} &    3.0\\
\hline
\multicolumn{7}{|c|}{$J^\pi=1/2^-$}\\
\hline
     2594 &    3.1 &   44.4 &    0.3 &    0.0 & {\bf \underline{52.0}} &    0.0\\
     2853 &    1.7 &   43.8 &    8.0 &    2.2 & {\bf \underline{44.2}} &    0.1\\
     2950 &    0.2 &    7.0 & {\bf \underline{85.4}} &    1.0 &    2.2 &    4.3\\
     3018 &   12.2 &   40.7 &    2.4 &    0.0 & {\bf \underline{44.3}} &    0.1\\
\hline
\multicolumn{7}{|c|}{$J^\pi=3/2^-$}\\
\hline
     2586 &    3.0 &   42.3 &    0.3 &    0.0 & {\bf \underline{54.3}} &    0.1\\
     2874 &    0.1 &   31.9 & {\bf \underline{42.1}} &    1.3 &   23.4 &    1.2\\
     2927 &    1.3 &   23.2 & {\bf \underline{53.9}} &    1.0 &   19.1 &    1.5\\
     3009 &   11.1 &   36.7 &    0.6 &    0.0 & {\bf \underline{51.3}} &    0.0\\
\hline
\end{tabular}
\end{center}
\caption{Masses (in MeV) and SU(8) configuration mixings (in \%) for the $\Lambda_c$ ground state and excited states with total angular momentum $J$ and parity $\pi$. The configuration is denoted by $^{2S+1}\mathrm{Flav}[\mathrm{Sym}]$ where $S$ is the total intrinsic spin of the quarks, Flav is the corresponding flavor multiplet and Sym the symmetry of the total spin-flavor SU(8) configuration. Dominant contributions are bold printed and underlined. We have omitted the generally very small negative energy contributions.}
\label{tab:su8confLamC}
\end{table}
\end{center}

Keeping the coupling $g_{nc}$ at zero, we increase the coupling $g_{nn}$ switching on the 't Hooft-interaction between the light quarks. We observe that in particular the lowest lying states $\Lambda_c\frac{1}{2}^+$, $\Lambda_c\frac{1}{2}^-$, $\Lambda_c\frac{3}{2}^-$ and the second lowest lying state $\Lambda_c\frac{1}{2}^+$ are very strongly influenced by 't Hooft's force. They are each lowered by approximately 230\,MeV. This can be understood when we analyze their spin-flavor SU(8) configurations listed in table \ref{tab:su8confLamC}. The aforesaid states are dominantly $^2 \bar 3_{\mathrm{M}}[\mathrm{S}]$ or $^2 \bar 3_{\mathrm{A}}[\mathrm{M}]$ configurations. The flavor states of the mixed anti-triplet $\bar 3_{\mathrm{M}}$ are anti-symmetric under the interchange of the two light quarks and the flavor states of the anti-symmetric anti-triplet $\bar 3_{\mathrm{A}}$ are anti-symmetric anyway. The total intrinsic spin of these states is not $\frac{3}{2}$ and thus the spin part not totally symmetric. Such spin-flavor combinations allow 't Hooft's force (which projects onto anti-symmetric spin and flavor pairs) to act unhindered which in fact it does.

\begin{center}
\begin{table}[t]
\begin{center}
\begin{tabular}{|c|rrrr|rr|}
\hline
Mass &
$\!\!{}^2      6_{\mathrm{M}}[\mathrm{S}]\!\!$&
$\!\!{}^2      6_{\mathrm{M}}[\mathrm{M}]\!\!$&
$\!\!{}^4      6_{\mathrm{M}}[\mathrm{M}]\!\!$&
$\!\!{}^2      6_{\mathrm{M}}[\mathrm{A}]\!\!$&
$\!\!{}^4      6_{\mathrm{S}}[\mathrm{S}]\!\!$&
$\!\!{}^2      6_{\mathrm{S}}[\mathrm{M}]\!\!$\\
\hline
\multicolumn{7}{|c|}{$J^\pi=1/2^+$}\\
\hline
     2459 & {\bf \underline{95.5}} &    3.1 &    0.0 &    0.0 &    0.0 &    1.3\\
     2947 & {\bf \underline{88.9}} &    6.3 &    0.0 &    0.1 &    0.0 &    4.5\\
     3048 &    0.0 &    0.1 &   36.1 &    0.8 & {\bf \underline{62.6}} &    0.1\\
     3088 &    1.6 & {\bf \underline{64.3}} &    0.0 &    0.1 &    0.2 &   33.5\\
\hline
\multicolumn{7}{|c|}{$J^\pi=3/2^+$}\\
\hline
     2539 &    0.0 &    0.0 &    3.0 &    0.0 & {\bf \underline{96.9}} &    0.0\\
     3010 &    0.4 &    0.1 &   19.3 &    0.5 & {\bf \underline{79.4}} &    0.2\\
     3031 & {\bf \underline{66.1}} &    4.3 &    1.2 &    0.7 &    1.9 &   25.7\\
     3049 &    1.3 &    4.3 &   35.9 &    1.4 & {\bf \underline{54.2}} &    2.7\\
\hline
\multicolumn{7}{|c|}{$J^\pi=1/2^-$}\\
\hline
     2769 &    3.4 & {\bf \underline{71.6}} &    2.7 &    0.2 &    0.0 &   22.1\\
     2817 &    0.1 &    3.6 & {\bf \underline{94.2}} &    0.0 &    1.9 &    0.1\\
     2929 &    0.1 &   20.5 &    1.2 &    2.4 &    0.0 & {\bf \underline{75.8}}\\
\hline
\multicolumn{7}{|c|}{$J^\pi=3/2^-$}\\
\hline
     2799 &    1.1 &   10.3 & {\bf \underline{59.0}} &    0.1 &    1.2 &   28.2\\
     2815 &    2.2 & {\bf \underline{50.1}} &   31.9 &    0.1 &    0.6 &   14.9\\
     2919 &    0.2 &   35.7 &    7.1 &    2.5 &    0.1 & {\bf \underline{54.2}}\\
\hline
\end{tabular}
\end{center}
\caption{Masses (in MeV) and SU(8) configuration mixings (in \%) for the $\Sigma_c$ ground state and excited states with total angular momentum $J$ and parity $\pi$.}
\label{tab:su8confSigC}
\end{table}
\end{center}

The opposite is true for the spin-flavor SU(8) configurations of the $\Sigma_c$ in table \ref{tab:su8confSigC}. All states belong to flavor 6-plets. The flavor states of the mixed sextet $6_{\mathrm{M}}$ are symmetric under the interchange of the two light quarks and the flavor states of the symmetric sextet $6_{\mathrm{S}}$ are symmetric anyway. This forbids 't Hooft's force to act between the $nn$-quark pairs which can also be seen in fig.~\ref{fig:charmed_tHooft01}.

Using the confinement potential alone, the calculated ground states $\Lambda_c\frac{1}{2}^+$ and $\Sigma_c\frac{1}{2}^+$ are degenerate. But the 't Hooft-interaction acting only on the $\Lambda_c\frac{1}{2}^+$ leads to their experimentally observed splitting, including the correct sign. This splitting amounts to 167\,MeV in experiment and is 239\,MeV when we only take into account 't Hooft's force acting between the light quarks. The latter number will be modified in the correct direction when we also include the 't Hooft-interaction between a light and the charm quark.

Looking again at fig.~\ref{fig:charmed_tHooft01} and the remaining states, we see that the experimental situation for the $\Lambda_c$ and $\Sigma_c$ resonances is already nicely explained by the confinement and the 't Hooft-potential between light quarks alone: There is a clear one-to-one correlation between experimental and theoretical states. Problematic remain the splittings between the lowest lying pairs $\Lambda_c\frac{1}{2}^-$, $\Lambda_c\frac{3}{2}^-$ and $\Sigma_c\frac{1}{2}^+$, $\Sigma_c\frac{3}{2}^+$. We note that the lowering of the former pair states is almost identically strong which spoils a direct explanation for their splitting. This is no surprise since they possess almost identical spin-flavor configurations, see table \ref{tab:su8confLamC}, but the lowering is too strong for the $\frac{3}{2}^-$-state. Nevertheless, we will see below that the splitting between the second pair can be impressively explained with the extension of 't Hooft's force to charm quarks.

Figure \ref{fig:charmed_tHooft02} shows on the left in each column the theoretical single charm baryon spectrum of the $\Xi_c$ resonances again determined by the confinement interaction alone which means that $g_{ns}=g_{nc}=g_{sc}=0$. Due to the quark flavor content $nsc$ there never is a contribution from the coupling parameterized by $g_{nn}$.

\begin{center}
\begin{table}[t]
\begin{center}
\begin{tabular}{|c|rrrr|rr|}
\hline
Mass &
$\!\!{}^2      6_{\mathrm{M}}[\mathrm{S}]\!\!$&
$\!\!{}^2      6_{\mathrm{M}}[\mathrm{M}]\!\!$&
$\!\!{}^4      6_{\mathrm{M}}[\mathrm{M}]\!\!$&
$\!\!{}^2      6_{\mathrm{M}}[\mathrm{A}]\!\!$&
$\!\!{}^4      6_{\mathrm{S}}[\mathrm{S}]\!\!$&
$\!\!{}^2      6_{\mathrm{S}}[\mathrm{M}]\!\!$\\
 &
$\!\!{}^2 \bar 3_{\mathrm{M}}[\mathrm{S}]\!\!$&
$\!\!{}^2 \bar 3_{\mathrm{M}}[\mathrm{M}]\!\!$&
$\!\!{}^4 \bar 3_{\mathrm{M}}[\mathrm{M}]\!\!$&
$\!\!{}^2 \bar 3_{\mathrm{M}}[\mathrm{A}]\!\!$&
$\!\!{}^4 \bar 3_{\mathrm{A}}[\mathrm{S}]\!\!$&
$\!\!{}^2 \bar 3_{\mathrm{A}}[\mathrm{M}]\!\!$\\
\hline
\multicolumn{7}{|c|}{$J^\pi=1/2^+$}\\
\hline
     2469 & 1.3 &    0.2 &    0.0 &    0.0 &    0.0 &    0.5\\
          & {\bf \underline{96.9}}&    0.6 &    0.0 &    0.0 &    0.0 &    0.3\\
     2595 & {\bf \underline{94.2}}&    2.5 &    0.0 &    0.0 &    0.0 &    1.0\\
          & 1.2 &    0.5 &    0.0 &    0.0 &    0.0 &    0.5\\
\hline
\multicolumn{7}{|c|}{$J^\pi=3/2^+$}\\
\hline
     2651 & 0.0 &    0.0 &    2.7 &    0.0 &   {\bf \underline{96.0}} &    0.0\\
          & 0.0 &    0.0 &    1.3 &    0.0 &    0.0 &    0.0\\
\hline
\multicolumn{7}{|c|}{$J^\pi=1/2^-$}\\
\hline
     2769 & 0.8 &   40.5 &    0.1 &    1.0 &    0.0 &    0.5\\
          & 0.0 &    0.0 &    0.0 &    0.0 &    0.0 &   {\bf \underline{57.1}}\\
\hline
\multicolumn{7}{|c|}{$J^\pi=3/2^-$}\\
\hline
     2771 & 0.7 &   44.5 &    0.2 &    0.9 &    0.0 &    0.4\\
          & 0.0 &    0.0 &    0.0 &    0.0 &    0.0 &   {\bf \underline{53.2}}\\
\hline
\end{tabular}
\end{center}
\caption{Masses (in MeV) and SU(8) configuration mixings (in \%) for the $\Xi_c$ ground state and excited states with total angular momentum $J$ and parity $\pi$.}
\label{tab:su8confXiC}
\end{table}
\end{center}

We see that only the $\Xi_c(2645)$ is really well described by the confinement potential alone. But increasing just the coupling $g_{ns}$ and keeping the other couplings $g_{nc}$ and $g_{sc}$ at zero, we observe that the two lowest lying degenerate states for the spin-parity combination $\frac{1}{2}^+$ separate creating the experimentally observed splitting between the $\Xi_c$ and the $\Xi'_c$. This behavior can be explained when we investigate their spin-flavor SU(8) configurations listed in table \ref{tab:su8confXiC}: They are almost entirely $^2 \bar 3_{\mathrm{M}}[\mathrm{S}]$ and $^2 6_{\mathrm{M}}[\mathrm{S}]$ configurations respectively. Like before 't Hooft's force acts on the flavor-$\bar 3$-plet but not on the flavor-6-plet members, so it lowers the $\Xi_c$ and leaves the $\Xi_c'$ virtually untouched. Their splitting is 108\,MeV experimentally and 159\,MeV when we consider the 't Hooft-interaction between the light quarks only. This theoretical number will also be improved when we include 't Hooft's interaction between all flavors.

We find the $\Xi(2645)$ as an almost pure $^4 6_{\mathrm{S}}[\mathrm{S}]$ configuration and thus a state entirely consisting of intrinsic spin $\frac{3}{2}$, see table \ref{tab:su8confXiC}. The 't Hooft-interaction will not influence this state, regardless between which quark flavors it acts.

The negative parity states $\Xi_c\frac{1}{2}^-$ and $\Xi_c\frac{3}{2}^-$ have practically the same configuration mixings and our residual interaction thus affects both states in nearly the same way, namely each state is lowered by 133\,MeV. But, just like for the $\Lambda_c\frac{3}{2}^-$, the lowering of the $\Xi_c\frac{3}{2}^-$ resonance is too strong to yield the correct experimental splitting between $\Xi_c(2790)$ and $\Xi_c(2815)$.

To sum up the situation for the $\Xi_c$ in fig.~\ref{fig:charmed_tHooft02}, it can be said that the gross experimental structure is already well described by the confinement potential and 't Hooft's residual interaction operating only between light quarks. However, the masses of the the important states $\Xi_c$ and $\Xi'_c$ are too large which will be cured below when including phenomenologically 't Hooft's force also operating between a light and a charm quark. 

We finally notice that for the $\Omega_c$ there is no 't Hooft-potential between two light quarks.

\subsection{Instanton effects between a light and a charm quark}

Based on the findings of the preceding section, we will now investigate the effects of the instanton-induced interaction when operating also between a light and the charm quark.

\begin{figure*}[p]
\begin{center}
\input{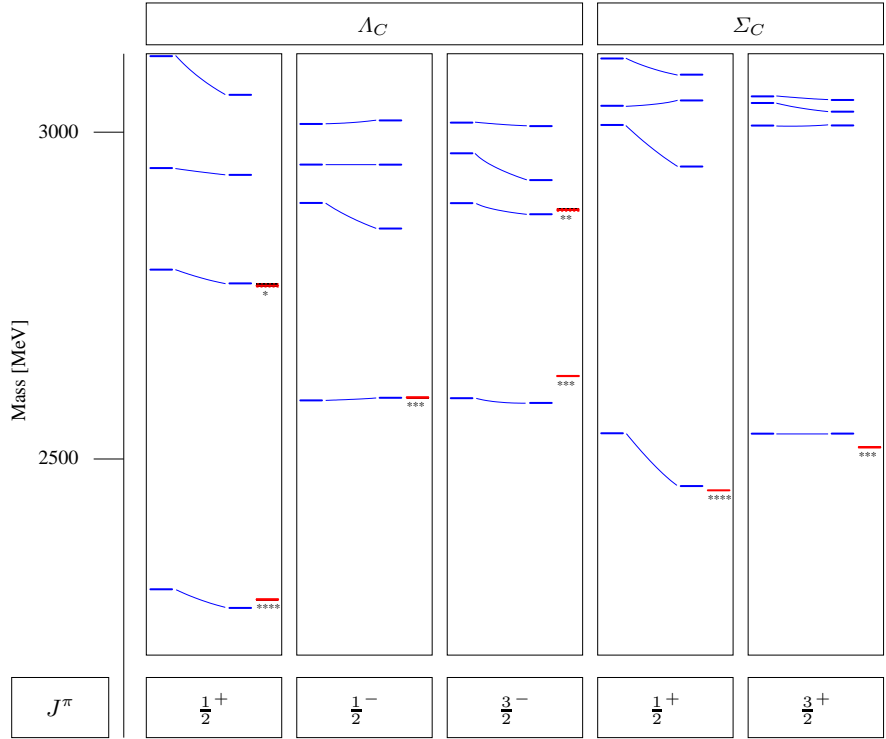}
\caption{The effects of the instanton-induced interaction between a light and the charm quark on the theoretical single charm baryon resonances $\Lambda_c$ and $\Sigma_c$ with total angular momentum $J$ and parity $\pi$. The spectrum on the left in each column is determined by the confinement and the residual interaction operating only between the light quarks. The following curves illustrate how the spectrum changes with increasing 't Hooft-coupling $g_{nc}$. The final value is shown in comparison with the experimental value shown on the right in each column.}
\label{fig:charmed_tHooft1}
\end{center}
\end{figure*}

Figure \ref{fig:charmed_tHooft1} shows on the left in each column the theoretical single charm baryon spectrum of the $\Lambda_c$ and $\Sigma_c$ resonances determined by the confinement interaction and the residual interaction operating only between light quarks, \textit{i.e.}~for $g_{nn}=136\,\mathrm{MeV\,fm}^3$ and $g_{nc}=0$.

\begin{figure*}[p]
\begin{center}
\input{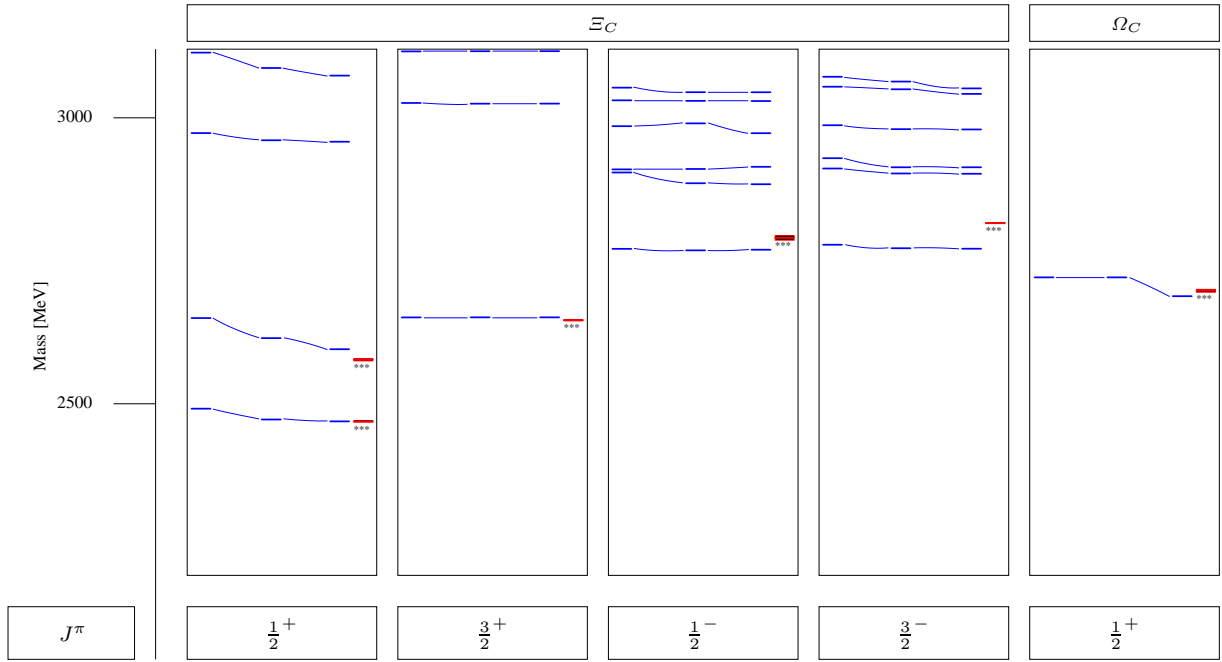}
\caption{The effects of the instanton-induced interaction between a light and the charm quark on the resonances $\Xi_c$ and $\Omega_c$. The spectrum on the left in each column is determined by the confinement and the residual interaction operating only between the light quarks. The curves illustrate how the spectrum changes with firstly increasing the 't Hooft-coupling $g_{nc}$ and secondly $g_{sc}$. The final value is shown in comparison with the experimental value.}
\label{fig:charmed_tHooft2}
\end{center}
\end{figure*}

Increasing the coupling $g_{nc}$, we mainly observe the lowering of the lowest lying $\Sigma_c\frac{1}{2}^+$-state by 80\,MeV describing excellently the mass of the $\Sigma(2455)$. This selective lowering also explains the mass splitting between the lowest lying $\Sigma_c\frac{1}{2}^+$- and $\Sigma_c\frac{3}{2}^+$-states which is again 80\,MeV and almost coincides with the experimental value of 66\,MeV. This behavior of 't Hooft's interaction can again be understood from inspection of the SU(8) spin-flavor configurations in table \ref{tab:su8confSigC}. The $\Sigma_c\frac{1}{2}^+$ is a $^26_{\mathrm{M}}[\mathrm{S}]$-state which can be influenced by the extended 't Hooft-force we are considering whereas the $\Sigma_c\frac{3}{2}^+$ is a $^46_{\mathrm{S}}[\mathrm{S}]$-state and thus not changed by 't Hooft's interaction for reasons we have already explained.

We also see a lowering of the $\Lambda_c$ whose position was 16\,MeV too high before and 13\,MeV too low after we have increased $g_{nc}$ to its maximum value. So we achieve also here a slight improvement by the additional residual interaction.

Due to this strong lowering of the theoretical $\Sigma_c(2455)$ we get in the Bethe-Salpeter-model a final splitting between the $\Lambda_c$ and the $\Sigma_c(2455)$ of 187\,MeV which is very close to the already mentioned experimental value of 167 MeV. The extended version of 't Hooft's force improves this number by 52\,MeV. 

The experimental splitting between the lowest lying $\Lambda_c\frac{1}{2}^-$- and $\Lambda_c\frac{3}{2}^-$-states remains unexplained. Although the additional 't Hooft-force acts only very weakly on these two states, the effect of it even has the wrong sign. Nevertheless, the mass of the $\Lambda_c(2593)$ is perfectly described.

We shall not forget the two- and one-star resonance whose masses are lowered by 17\,MeV and 21\,MeV respectively, almost matching their alleged experimental counterparts.

Figure \ref{fig:charmed_tHooft2} shows on the left in each column the theoretical single charm baryon spectrum of the $\Xi_c$ and $\Omega_c$ resonances determined by the confinement interaction and the residual interaction operating only between the light quarks which means that $g_{ns}=94.0\,\mathrm{MeV\,fm}^3$ and $g_{nc}=g_{sc}=0$. We first increase the coupling $g_{nc}$ to its maximum value and depict the resulting masses in the middle of each column. We then increase the coupling $g_{sc}$ and compare the final theoretical values to the experimental ones.

We especially observe further lowerings of the theoretical $\Xi_c$ and $\Xi'_c$ which can now happen to both states, see again table \ref{tab:su8confXiC}, in contrast to the situation in the previous section where we have considered 't Hooft's interaction only between light flavor quarks. 't Hooft's interaction between the quark pairs $nc$ and $sc$ lowers the position of the $\Xi_c$ successively by 19 and 3\,MeV and the position of the $\Xi_c'$ by 35 and 20\,MeV reproducing perfectly the mass of the $\Xi_c$ and describing very well the mass of the $\Xi_c'$. Moreover, the $\Xi_c'$ is lowered by a larger amount than the $\Xi_c$ which improves the theoretical mass splitting between them by 16\,MeV to 143\,MeV. This is closer to the experimental splitting of 108\,MeV than without the phenomenological extension of 't Hooft's force to charm quarks.

Due to its SU(8) spin-flavor configuration the theoretical mass for the $\Xi_c(2645)$ is never influenced by 't Hooft's force which is reasonable because it is already well described by the confinement potential alone.

Concerning the theoretical states for the $\Xi_c(2790)$ and the $\Xi_c(2815)$, the circumstances are very similar to the case of the $\Lambda_c(2593)$ and the $\Lambda_c(2625)$. These pairs not only have exactly the same quantum numbers but show the same behavior when compared to the experimental states. The effects of 't Hooft's force are almost negligible but again have the wrong sign and cannot explain the experimentally observed splitting.

\begin{center}
\begin{table}[t]
\begin{center}
\begin{tabular}{|c|rrrr|rr|}
\hline
Mass &
$\!\!{}^2      6_{\mathrm{M}}[\mathrm{S}]\!\!$&
$\!\!{}^2      6_{\mathrm{M}}[\mathrm{M}]\!\!$&
$\!\!{}^4      6_{\mathrm{M}}[\mathrm{M}]\!\!$&
$\!\!{}^2      6_{\mathrm{M}}[\mathrm{A}]\!\!$&
$\!\!{}^4      6_{\mathrm{S}}[\mathrm{S}]\!\!$&
$\!\!{}^2      6_{\mathrm{S}}[\mathrm{M}]\!\!$\\
\hline
\multicolumn{7}{|c|}{$J^\pi=1/2^+$}\\
\hline
     2688 & {\bf \underline{97.5}} &    1.5 &    0.0 &    0.0 &    0.0 &    1.0\\
     3169 & {\bf \underline{84.3}} &    8.1 &    0.0 &    0.1 &    0.0 &    7.3\\
\hline
\multicolumn{7}{|c|}{$J^\pi=3/2^+$}\\
\hline
     2721 &    0.0 &    0.0 &    2.0 &    0.0 & {\bf \underline{97.9}} &    0.0\\
\hline
\end{tabular}
\end{center}
\caption{Masses (in MeV) and SU(8) configuration mixings (in \%) for the $\Omega_c$ ground state and excited states with total angular momentum $J$ and parity $\pi$.}
\label{tab:su8confOmeC}
\end{table}
\end{center}

When we only take the confinement potential into account the theoretical position of the $\Omega_c$ is overestimated by 23\,MeV compared to the experiment. There is trivially no contribution from 't Hooft's force between a quark pair $nc$ but a strong lowering by 33\,MeV due to 't Hooft's force between the quark pair $sc$ although the value for $g_{sc}$ is relatively small. Looking at its SU(8) spin-flavor configuration in table \ref{tab:su8confOmeC}, the $\Omega_c$ is indeed almost a pure $^26_{\mathrm{M}}[\mathrm{S}]$-state. With 't Hooft's force switched on we underestimate the mass of the $\Omega_c$ only by 10\,MeV which is much better than without 't Hooft's force.

Summarizing the effects of the 't Hooft-interaction, we can clearly state that the phenomenological extension of the instanton-in\-duced force to charmed baryons leads to visible improvements in the whole fine structure of the single charm baryon spectrum.

\subsection{One-gluon-exchange}

For comparison we replace within the same framework our phenomenologically motivated extension of 't Hooft's force by the instantaneous version of the one-gluon-exchange as a residual interaction operating between a light and a heavy quark. The interaction between the light quarks, which is taken to be the instanton-induced force, remains the same and forms the basis for all calculations presented in this work. Its effects on the single charm baryon spectrum is already discussed in sec.~\ref{sec:instlight}. 

In the previous section we have seen that the extended 't Hooft-interaction is able to improve the fine structure of the spectrum which is based on the confinement potential and the 't Hooft-interaction between the light quarks. We now show that the one-gluon-exchange between a light and a heavy quark is capable of explaining some features of the experimental spectrum but cannot account for some distinguished properties concerning the fine structure. 

The confinement potential used here primarily models the long-ranged confining forces. But due to its negative offset it also induces a short-ranged attractive influence which acts between all quark pairs, regardless if they are light or heavy. When we want to consider the one-gluon-exchange and 't Hooft's interaction combined in the same framework but acting between differently flavored quark pairs, we have to take into account that the one-gluon-exchange also has a short-ranged attractive effect. Avoiding a double-counting of these attractive short-ranged forces, we therefore neglect the Coulomb-term $D_{00}$ in eq.~(\ref{eqn:OGEpot}). 

\begin{figure*}[t]
\begin{center}
\input{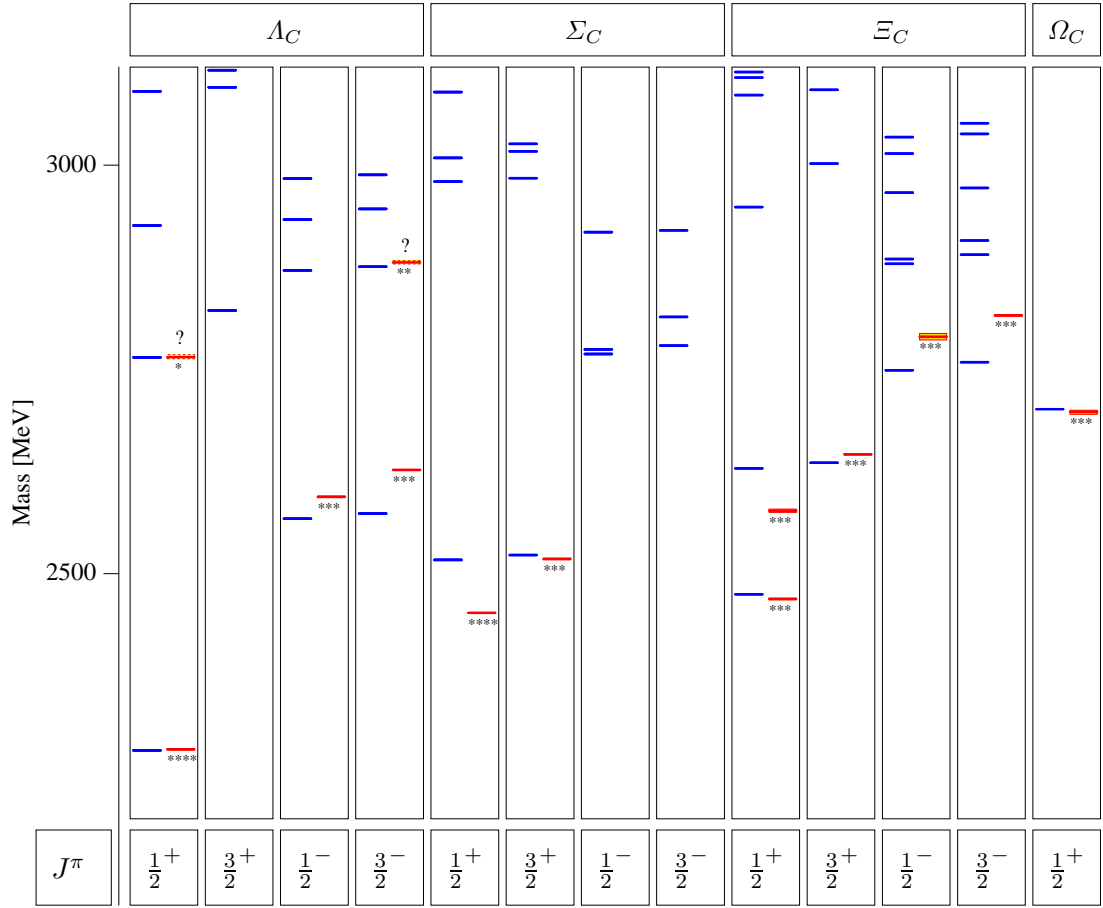}
\caption{The single charm baryon spectrum calculated in the Bethe-Salpeter-model using an appropriate three-quark confinement potential, the one-gluon-exchange between a light and a heavy quark and 't Hooft's force between the remaining flavor combinations (on the left side of each column) in comparison to the experimental situation taken from \cite{Eidelman:2004wy} (on the right side where the status is indicated by stars, the mass uncertainty by a shaded box and the lack of knowledge of quantum numbers by a question mark) in dependence of the total angular momentum $J$ and the parity $\pi$.}
\label{fig:onegluonexchange}
\end{center}
\end{figure*}

Figure \ref{fig:onegluonexchange} shows the resulting single charm baryon spectrum where the new parameters, namely the strong coupling constant $\alpha_s=1.0$ and the charm constituent quark mass $m_c=1940\,$MeV, are fitted simultaneously to the three- and four-star charmed baryons.

We find that there is a clear one-to-one correlation between experimental and theoretical states and in particular the masses of the lowest lying states $\Lambda_c\frac{1}{2}^+$, $\Sigma_c\frac{3}{2}^+$, $\Xi_c\frac{1}{2}^+$, $\Xi_c\frac{3}{2}^+$ and $\Omega_c\frac{1}{2}^+$ are very well described.

But the one-gluon-exchange fails completely in explaining the positions of the $\Sigma_c(2455)$ and the $\Xi'_c$, missing them by 65\,MeV and 52\,MeV. Remember that both states could be accurately described by 't Hooft's force.

The one-gluon-exchange also underestimates the resonance masses of the problematic pairs $\Lambda_c(2593)$, $\Lambda_c(2625)$ and $\Xi_c(2790)$, $\Xi_c(2815)$, on average by 20\,MeV more for each state when compared to 't Hooft's force. But the one-gluon-exchange can allusively account for their splittings: They are 7\,MeV and 10\,MeV respectively and are approximately by factors five and three too small when compared to the experimental splittings of 33\,MeV and 26\,MeV.

We can summarize that the one-gluon-exchange is able to describe the single charm baryon spectrum but the splittings it generates (in spite of the embarrassing large coupling $\alpha_s$) are too small to explain the fine structure satisfactorily. We obtain overall better results using the extended version of 't Hooft's interaction.

\subsection{Double and triple charm baryons}

\begin{center}
\begin{table}[t]
\begin{center}
\begin{tabular}{|c|c|c|c|}
\hline
State&$J^\pi$&BSM&BSM\\
&&conf.&full\\
\hline
$\Xi_{cc}$    &$1/2^+$&3723&3642\\
$\Xi_{cc}$    &$3/2^+$&3723&3723\\ 
$\Xi_{cc}$    &$1/2^-$&3941&3920\\
$\Xi_{cc}$    &$3/2^-$&3941&3920\\ 
$\Omega_{cc}$ &$1/2^+$&3765&3732\\
$\Omega_{cc}$ &$3/2^+$&3765&3765\\ 
$\Omega_{cc}$ &$1/2^-$&3994&3986\\
$\Omega_{cc}$ &$3/2^-$&3994&3986\\
$\Omega_{ccc}$&$1/2^+$&5216&5216\\
$\Omega_{ccc}$&$3/2^+$&4773&4773\\
$\Omega_{ccc}$&$1/2^-$&5019&5019\\
$\Omega_{ccc}$&$3/2^-$&5014&5014\\
\hline
\end{tabular}
\end{center}
\caption{The theoretical mass positions (in MeV) of the lowest lying double and triple charm baryons in dependence of the total angular momentum $J$ and parity $\pi$ calculated within the Bethe-Salpeter-model (BSM) using an appropriate three-quark confinement potential and 't Hooft's force between a light and a charm quark (labeled with ``full''). We also show the computed masses using the confinement potential only (labeled with ``conf.'').}
\label{tab:doublytriplycharmedmasses}
\end{table}
\end{center}

Without introducing additional free parameters we also compute the masses of double and triple charm baryons. Encouraged by its success in describing the fine structure of the single charm baryon spectrum, we again use the extended 't Hooft-force between a light and a charm quark which now is the only residual interaction since there is only one light quark. Table \ref{tab:doublytriplycharmedmasses} shows the results for the lowest lying states of the $\Xi_{cc}$, $\Omega_{cc}$ and $\Omega_{ccc}$. To demonstrate the effects of 't Hooft's force we also present the results using the confinement potential alone.

The only experimental clue for a double charm bary\-on, namely the $\Xi_{cc}$, stems from the SELEX Collaboration at Fermilab \cite{Mattson:2002vu}. The mass of the $\Xi_{cc}$ is supposedly $(3519\pm 1)\,$MeV, its quantum numbers are unknown. The Particle Data Group Collaboration does not consider this state as established and gives it only one star \cite{Eidelman:2004wy}.

Taking this experimental finding seriously, all our theoretical predictions given in table \ref{tab:doublytriplycharmedmasses} are too high. The most obvious identification is with the lowest lying state which has the spin-parity combination $\frac{1}{2}^+$ and mass 3642\,MeV. The deviation to the experiment is thus 3.5\% which is still not too bad.
 
Various other models predict the lowest lying $\Xi_{cc}\frac{1}{2}^+$-state also too high and very comparable to our result: In the framework of a relativistic quasi potential quark model \cite{Ebert:1996ec} one obtains 3660\,MeV and in a similar and more recent calculation by the same authors \cite{Ebert:2002ig} 3620\,MeV, in the framework of a simple potential model \cite{Richard:1994ae} one obtains 3630\,MeV, by exploiting regularities in the hadron interaction energies to obtain sum rules \cite{Lichtenberg:1995kg} one obtains 3676\,MeV and in quenched lattice calculations \cite{Woloshyn:2000fe,Mathur:2002ce} one obtains approximately 3600\,MeV.

Looking again at table \ref{tab:doublytriplycharmedmasses}, we see that 't Hooft's force lowers the $\Xi_{cc}\frac{1}{2}^+$-state by the relatively large amount of 81\,MeV but it seems that its strength is somehow not strong enough to explain the small experimental value.

\vspace{20pt}

\section{Conclusion} \label{sec:conclusion}

In the framework of a relativistically covariant constituent quark model we calculated on the basis of the Bethe-Salpeter-equation in its instantaneous approximation mass spectra of single, double and triple charm baryons. We used a linearly rising three-body confinement potential and as a residual interaction between light quarks always 't Hooft's instanton-induced force. Both interactions were already fixed by the non-strange and strange baryon spectrum and were left unchanged in the calculations of this work. The Bethe-Salpeter-model was therewith able to give an adequate description of the whole non-strange and strange baryon spectra.

In this paper we have extended phenomenologically 't Hooft's force to include also charm quarks introducing two additional new free coupling parameters. However, the foundations of such a force are by no means clear because strictly it exist only for nearly massless quarks. Our investigation is therefore to be considered to be of exploratory nature. Nevertheless, we keep for simplicity the name instanton-induced force. We showed the computed spectrum of the single charm baryons, compared it to the experiment and discussed in detail the instanton-induced effects. We have found not only a very good overall agreement between theory and experiment but also the hyperfine splittings were accurately described. The extended version of 't Hooft's force can in particular correctly explain the experimental mass splittings between the $\Lambda_c$ and the $\Sigma_c(2455)$ and the $\Xi_c$ and the $\Xi_c'$.

Within the same framework we have considered an alternative effective interaction by replacing the 't Hooft-interaction between a light and the charm quark by the one-gluon-exchange. We reject the one-gluon-exchange phenomenologically due to its insufficiency to reproduce the correct fine structure of the charmed baryon spectrum.

Finally we have given many predictions for masses of double and triple charm baryons to be confronted with results from further experimental analyses.

Our theoretical investigations are still incomplete because so far we have not computed electroweak couplings and strong decays of charmed baryons. This is work in progress. Indeed, a number of experimental results are now available which could be used to further test our specific model. With the exception of $\Lambda_c$, semileptonic decays have been computed so far only for light flavored baryons \cite{Migura:2006en}.

\section*{Acknowledgments}
Financial support from the Deutsche Forschungsgemeinschaft by the
SFB/Transregio 16 ``Subnuclear Structure of Matter'' and from the European
Community-Research Infrastructure Activity under the FP6 "Structuring the
European Research Area" programme (HadronPhysics, contract number
RII3-CT-2004-506078) is gratefully acknowledged.

\end{document}